\documentclass[aps, prd, twocolumn, superscriptaddress, nofootinbib, amsmath, amssymb]{revtex4-2}

\usepackage{graphicx}  
\usepackage{mathrsfs}
\usepackage{amsmath}
\usepackage{bbold} 
\usepackage{braket}
\usepackage[normalem]{ulem}
\usepackage[breaklinks=true,colorlinks,citecolor=blue,linkcolor=blue,urlcolor=blue]{hyperref}

\usepackage{xargs} 
\usepackage[colorinlistoftodos,prependcaption,textsize=tiny]{todonotes}

\makeatletter
\newcommand{\fmarki}{*}
\newcommand{\fmarkii}{\ensuremath{\dagger}}
\newcommand{\fmarkiii}{\ensuremath{\ddagger}}
\newcommand{\fmarkiv}{\ensuremath{\mathsection}}
\newcommand{\fmarkv}{\ensuremath{\mathparagraph}}
\newcommand{\fmarkvi}{\ensuremath{\|}}
\newcommand{\fmarkvii}{**}
\newcommand{\fmarkviii}{\ensuremath{\dagger\dagger}}
\newcommand{\fmarkix}{\ensuremath{\ddagger\ddagger}}
                
\def\@fnsymbol#1{{\ifcase#1\or \fmarki\or \fmarkii\or \fmarkiii\or \fmarkiv\or \fmarkv\or \fmarkvi\or \fmarkvii\or \fmarkviii\or \fmarkix \else\@ctrerr\fi}}
\makeatother

\renewcommand{\fmarkix}{\ensuremath\ddag\ddag}
\def\@fnsymbol#1{{\ifcase#1\or \fmarki\or \fmarkii\or \fmarkiii\or \fmarkiv\or \fmarkv\or \fmarkvi\or \fmarkvii\or \fmarkviii\or \fmarkix \else\@ctrerr\fi}}
\makeatother

\newcommand{\Id}{\mathbb{1}}
\DeclareMathOperator{\Had}{Had}

\begin{document}

\title{Quantum Algorithms for the computation of quantum thermal averages at work}

\author{Riccardo~Aiudi}
\email{riccardo.aiudi@unipr.it}
\affiliation{Dipartimento di Fisica, Universit\`a di Parma, Parco Area delle Scienze 7 /A, 43100 Parma, Italy}

\author{Claudio~Bonanno}
\email{claudio.bonanno@csic.es}
\affiliation{Instituto de F\'isica Te\'orica UAM-CSIC, c/ Nicol\'as Cabrera 13-15, Universidad Aut\'onoma de Madrid, Cantoblanco, E-28049 Madrid, Spain}

\author{Claudio~Bonati}
\email{claudio.bonati@unipi.it}
\affiliation{Dipartimento di Fisica dell'Universit\`a di Pisa and INFN --- Sezione di Pisa, Largo Pontecorvo 3, I-56127 Pisa, Italy.}

\author{Giuseppe~Clemente}
\email{giuseppe.clemente@desy.de}
\affiliation{Deutsches Elektronen-Synchrotron (DESY), Platanenallee 6, 15738
Zeuthen, Germany}

\author{Massimo~D'Elia}
\email{massimo.delia@unipi.it}
\affiliation{Dipartimento di Fisica dell'Universit\`a di Pisa and INFN --- Sezione di Pisa, Largo Pontecorvo 3, I-56127 Pisa, Italy.}

\author{Lorenzo~Maio}
\email{lorenzo.maio@phd.unipi.it}
\affiliation{Dipartimento di Fisica dell'Universit\`a di Pisa and INFN --- Sezione di Pisa, Largo Pontecorvo 3, I-56127 Pisa, Italy.}

\author{Davide~Rossini}
\email{davide.rossini@unipi.it}
\affiliation{Dipartimento di Fisica dell'Universit\`a di Pisa and INFN --- Sezione di Pisa, Largo Pontecorvo 3, I-56127 Pisa, Italy.}

\author{Salvatore~Tirone}
\email{salvatore.tirone@sns.it}
\affiliation{Scuola Normale Superiore, I-56126 Pisa, Italy}

\author{Kevin~Zambello}
\email{kevin.zambello@pi.infn.it}
\affiliation{Dipartimento di Fisica dell'Universit\`a di Pisa and INFN --- Sezione di Pisa, Largo Pontecorvo 3, I-56127 Pisa, Italy.}

\begin{abstract}
Recently, a variety of quantum algorithms have been devised to estimate thermal averages 
on a genuine quantum processor. In this paper, we consider the practical implementation
of the so-called Quantum-Quantum Metropolis algorithm. As a testbed for this purpose, 
we simulate a basic system of three frustrated quantum spins and discuss its systematics,
also in comparison with the Quantum Metropolis Sampling algorithm.

\end{abstract}

\maketitle

\section{Introduction}\label{sec:intro}

The advent of quantum computation is expected to lead to breakthroughs in
various fields of computational science~\cite{Boixo:2018, Arute:2019, Zhu:2022}. 
In fact, it may disclose novel pathways to the solution of notable unsolved questions lying at the basis of classically intractable problems, from quantum chemistry, to condensed-matter and high-energy physics~\cite{QChem_rev_2020, NISQ_rev_2022}.
One of such examples deals with the physics of fundamental interactions, in particular 
when considering the strongly coupled regime, which is not treatable by 
perturbative analytical tools.  Classical computational schemes, based on a 
discretized path integral formulation, are indeed known to face hard and yet 
unsolved difficulties. This happens, for instance, when considering real-time 
processes and non-equilibrium physics, or even in the equilibrium case when the 
path-integral measure is not positive defined (as it happens at finite baryon density), 
a fact which prevents the application of classical Monte Carlo algorithms. Such an
algorithmic obstruction is, for example, the main reason for our incomplete
knowledge of the QCD phase diagram and of the physics of strongly interacting
matter at finite density, which is required for the investigation of compact
astrophysical objects~\cite{Shapiro:1983, Rajagopal:2001}.

In the case of equilibrium physics, one needs
to devise quantum algorithms capable to efficiently explore the Gibbs
ensemble of the target quantum system.  At present, the availability of quantum
resources adequate for the numerical investigation of systems of direct
physical interest, such as QCD, is still far from being achieved. Nevertheless,
a variety of candidate quantum algorithms have been already proposed, either by 
directly computing observables on the (mixed) thermal state~\cite{Poulin_2009, Bilgin_2010, Riera_2012, Wu_2019, Zhu_2020}, 
or by preparing an ensemble of pure states, sampled with the proper thermal 
distribution~\cite{QMS_paper, QQMA_paper, Somma_Gibbs_Sampling, Moussa_2019, Motta_2020, Sun_2021, Lu_2021, Yamamoto:2022jes,
Selisko_2022, Davoudi:2022uzo, Ball_2022, Powers_2023, Fromm_2023}. At the present stage, it is thus important
to investigate how the practical implementation of such algorithms works in simplified models, 
in order to better understand their systematics and pave the way to 
future and more realistic applications.

In Ref.~\cite{Qubipf_QMS}, some of the present authors have already focused on the so-called Quantum Metropolis Sampling (QMS) 
algorithm~\cite{QMS_paper}, applied to a simple frustrated system made up of 
three quantum spins, which presents a sign problem when formulated in the 
path-integral approach. The QMS algorithm is based on a quantum Metropolis step, 
by which one can implement a quantum Markov chain across the Hamiltonian eigenstates 
of the system, which is capable of correctly sampling, after a proper thermalization 
time, the quantum eigenstates with the desired ensemble probability.
In this paper, we also consider the alternative Quantum-Quantum Metropolis Algorithm 
(Q$^2$MA)~\cite{QQMA_paper}, which is based on a quite different strategy. In a 
few words, the idea is to search for a pure quantum state, the so-called 
\emph{coherent encoding of the thermal state} (CETS), with the property that 
a measurement of the Hamiltonian on such state returns a given eigenstate with 
the correct Gibbs distribution.
The search follows a Grover-like quantum approach, 
hence the double ``Quantum'' in the name of the algorithm, which therefore, at least 
in principle, promises a more effective quantum advantage. 

A fundamental ingredient for both approaches is the Quantum Phase Estimation (QPE)
algorithm~\cite{NielsenChuang:2010, Benenti:2018, Kitaev_1995, Cleve_1998}, which allows to estimate the 
energy eigenvalues once the Hamiltonian has been properly encoded in the quantum computer.
However, the overall strategy used in the two methods is quite different: in
the QMS algorithm the distribution is sampled via the Metropolis step, 
while in the Q$^2$MA algorithm it is encoded in the superposition amplitudes
of a pure state, which contains the information about the finite-temperature 
density matrix of the system, once the auxiliary registers are traced out.  

We are not aware of practical implementations and benchmarks of the Q$^2$MA algorithm 
in concrete examples, and the main purpose of the present study is to fill this gap, by 
considering the same target system of Ref.~\cite{Qubipf_QMS}. As we will discuss 
in more details later on, there are some aspects of the algorithm which make its 
practical implementation nontrivial. 

A different issue regards the numerical efficiency, measured in terms of the 
number of quantum gates needed to reach a given uncertainty on the final determination 
of the quantum thermal averages, for which we present a preliminary comparison between
the two algorithms. In carrying out such an analysis, particular attention
is taken towards both statistical and systematic contributions 
to the final error budget. 

The paper is organized as follows. In Sec.~\ref{sec:algo}, we first review
the basic features of the QMS and the Q$^2$MA algorithms, and then comment on
the possible sources of systematical errors. In Sec.~\ref{sec:model},
we introduce the specific quantum spin system tested here and all the metrics 
used to benchmark the systematical errors. In Sec.~\ref{sec:numres}, we present 
the results of our investigation, first assuming an exact encoding of the energy levels 
and then relaxing such constraint. Finally, our conclusions are drawn in Sec.~\ref{sec:conclusions}.

\section{Algorithms}\label{sec:algo}

The main focus of the present study is on the Q$^2$MA algorithm, since a practical 
implementation of the QMS algorithm has been already presented and discussed in 
Ref.~\cite{Qubipf_QMS}. However, we find it useful to give a brief overview of 
both algorithms, in order to clarify their computational requirements and to better 
identify the possible sources of systematical errors in both cases.

\subsection{Quantum Metropolis Sampling}\label{subsec:qms}

The QMS algorithm follows quite closely the scheme of the classical Metropolis algorithm 
(see Refs.~\cite{QMS_paper, Qubipf_QMS} for further details): a Markov chain is built in 
such a way that, at each step, one gets an eigenstate $|\varphi_k\rangle$ of the Hamiltonian 
of the system under study, with eigenvalue $E_k$. This is selected with a probability given 
(asymptotically after thermalization) by its Boltzmann weight $e^{-\beta E_k}/Z(\beta)$, where
$\beta=1/(k_B T)$ is the inverse temperature and $Z(\beta)$ is the partition function.

Four registers are required: a register encoding the state of the system 
(denoted by the subscript $1$), two energy registers 
encoding the energies before (denoted by $2$) and after (denoted by $3$)
the Metropolis step, if accepted, and finally a single-qubit acceptance register (denoted by $4$). 
The layout of the quantum state needed by the QMS algorithm is thus the following:
\begin{equation}
\ket{\text{acc}}_4 \ket{E_{\text{new}}}_3 \ket{E_{\text{old}}}_2 \ket{\mathrm{state}}_1\ .
\end{equation}

The first step of the Markov chain is the initialization of the system state in
register 1 to an arbitrary eigenstate of the Hamiltonian (and of register 2
to the corresponding eigenvalue). If the quantum state of the QMS algorithm has
been prepared in the initial state $\ket{0}_4\ket{0}_3\ket{0}_2\ket{0}_1$, 
this can be realized by using a QPE between registers $1$ and $2$, 
followed by a measure of register $2$:
\begin{equation}
\begin{aligned}
\ket{0}_2 \ket{0}_1 &\xrightarrow{\mathrm{QPE}_{1,2}} 
\sum_{k^\prime} \alpha_{k^\prime} \ket{E_{k^\prime}}_2 \ket{\varphi_{k^\prime}}_1 \\
 &\xrightarrow{\mathrm{Meas}_{2}} \ket{E_{k}}_2 \ket{\varphi_{k}}_1 \ .
\end{aligned}
\end{equation}
Registers $3$ and $4$ stay unmodified in this initial step.

A single step of the Markov chain involves an appropriate generalization of the
Metropolis accept/reject algorithm~\cite{Metropolis:1953am} to the quantum
case. In order to update the state, we apply to register 1 a unitary
operator $C$ randomly selected from a set $\mathcal{C}$; this set has to be
large enough to ensure mixing between all eigenstates (ergodicity) and that if
$A\in \mathcal{C}$ then also $A^{-1}\in\mathcal{C}$ (reversibility). Apart from
these general requirements, there is still much freedom in the choice of the
operators entering the set $\mathcal{C}$, freedom that can eventually be used to
optimize the algorithm. Thus
\begin{align}
\ket{\varphi_k}_1 \xrightarrow{C\in\mathcal{C}} \sum\limits_p x_{k,p}^{(C)} \ket{\varphi_p}_1\ .
\end{align}
At this point, we perform a second QPE between the register of the system
(labeled as $1$) and the new energy register (labeled as $3$):
\begin{equation}
\begin{aligned}
    &\sum\limits_p x_{k,p}^{(C)} \ket{0}_4\ket{0}_3\ket{E_k}_2 \ket{\varphi_p}_1 \\
    &\xrightarrow{\mathrm{QPE}_{1,3}} 
    \sum\limits_p x_{k,p}^{(C)} \ket{0}_4\ket{E_p}_3\ket{E_k}_2 \ket{\varphi_p}_1\ .
\end{aligned}
\end{equation}

To introduce the information about the Boltzmann weights, one applies an oracle operator 
$G$ which reads out the difference between the two energy registers and acts on the 
acceptance qubit as follows:
\begin{equation}
\begin{aligned}
  & \ket{0}_4 \ket{E_p}_3 \ket{E_k}_2 \ket{\psi_p}_1 \xrightarrow{G}  \\
  & \left( \sqrt{f_{p,k}}\ket{1}_4 + \sqrt{1-f_{p,k}} \ket{0}_4  \right)\otimes \ket{E_p}_3 \ket{E_k}_2 \ket{\psi_p}_1\, ,
\end{aligned}
\label{eq:metro_filter_rotation}
\end{equation}
where
\begin{equation}
  f_{p,k} = \min\left(1, e^{-\beta (E_p - E_k)}\right),
\end{equation}
is the usual Metropolis acceptance probability.  At this stage, one performs a
measurement in the acceptance register 4, which can only take two outcomes: the value $1$
with probability $\sum_p  |x_{k,p}^{(C)}|^2 f_{p, k}$ and the value $0$ with the 
complementary probability.  The first case corresponds to the ``accepted''
move, and the resulting state will be a superposition of eigenstates of the
form
\begin{align}
    \sum\limits_p x_{k,p}^{(C)}\sqrt{f_{p,k}} \ket{1}_4\ket{E_p}_3\ket{E_k}_2 \ket{\varphi_p}_1\ .
\end{align}
The new configuration of the Markov chain (and thus the corresponding
eigenstate in register $1$) is obtained by measuring the new energy register
$3$. From this configuration, the update procedure can be iterated by applying a
new randomly chosen unitary operator from the set $\mathcal{C}$.  
If, on the other hand, the outcome of the measure on the acceptance register 
is $0$, one needs to revert
the system to an eigenstate with the same energy as the one that was previously
present\footnote{There is no need for the system state to be reverted exactly
to the same state that was present before the application of the unitary
operator $C$, because this kind of rejection can be considered as a
microcanonical step in the classical Metropolis algorithm.}. This reversal
operation can be performed by applying backward the previous sequence of
unitary operators (i.e., QPE and $C$), followed by a measurement on the energy
register until the energy measured matches $E_k$ (see
Refs.~\cite{QMS_paper,Qubipf_QMS} for further details).

The above description of the QMS follows the original paper~\cite{QMS_paper}, 
where two different energy registers were used. We should however stress that 
it is possible to use just a single energy register:
since $E_{\text{old}}$ is measured only at the beginning of each MC step,  
its value can be stored in a classical register to be later used by the oracle 
$G$ in Eq.~\eqref{eq:metro_filter_rotation}, thus halving the number of qubits 
needed to represent the energies.

\subsection{Quantum-Quantum Metropolis Algorithm}\label{subsec:qqma}

The goal of the Q$^2$MA~\cite{QQMA_paper}, is to build a CETS containing the whole information on the Gibbs ensemble 
in the entanglement between the quantum registers. Its explicit form can be 
written as
\begin{equation}\label{eq:qqma_cets}
|\alpha_0 (\beta)\rangle = \sum_i \sqrt{e^{-\beta E_i} / Z(\beta)}\ket{\varphi_i}
\ket{\tilde{\varphi}_i}\ket{0}\ , 
\end{equation}
where $\ket{\varphi_i}$ denotes, as in the previous section, the Hamiltonian
eigenstate with eigenvalue $E_i$, $\ket{\tilde{\varphi_i}}$ is its complex
conjugate copy, while $\ket{0}$ is an ancillary register needed for an operation
analogous to the one of Eq.~\eqref{eq:metro_filter_rotation}. The presence of 
the complex conjugate copy of the system state has a double reason: building 
the system's density matrix and computing energy differences through a QPE routine.
The reason for the suffix $0$ of $\alpha_0$ will become clear in the following.
The state $\ket{\alpha_0(\beta)}$ is (apart from the irrelevant
ancilla $\ket{0}$) the purified form of the system's density matrix 
\begin{equation}\label{eq:rho_th}
\rho(\beta)=\frac{1}{Z(\beta)} \sum_i e^{-\beta E_i} \ket{\varphi_i} \bra{\varphi_i}\ .
\end{equation}

The core of the Q$^2$MA algorithm resides in the construction 
of the so-called \textit{generalized Szegedy operator} $W$, 
which is described in Ref.~\cite{QQMA_paper}. A fundamental ingredient for this 
construction is the so called ``kick'' operator $K$, which is a unitary operator
symmetric in the computational basis. The matrix elements of $K$ between eigenstates 
of the quantum Hamiltonian correspond to an ``a priori'' selection probability that, 
together with the Metropolis filter, can be used to sample the energy eigenstates 
by a Markov chain which has the CETS as invariant distribution 
(corresponding to the eigenvalue 1 of the Markov chain stochastic matrix). This procedure is 
very general, however, until an operator $K$ is specified for a given quantum Hamiltonian, 
it can not be shown that the corresponding Markov chain is ergodic, and thus that the CETS 
is the \emph{unique} eigenvector with eigenvalue 1 of 
the Szegedy operator. Let us assume for the moment that this is the case; we will come back 
to this point at the end of this section.

At this point, it is important to recall what are the main conceptual differences between 
the QMS and the Q$^2$MA algorithm, which have been already illustrated in 
Ref.~\cite{QQMA_paper}. The QMS is a quantum algorithm in the sense that its 
purpose is to exploit a quantum computing device to implement a Markov chain within 
the Hilbert space of a given quantum system. Apart from this, the main conceptual 
scheme is that of classical Markov chains, with additional limitations related
to the no-cloning theorem. On the other hand, the conceptual scheme of the Q$^2$MA is 
typical of quantum searching algorithms attaining a quadratic advantage, the 
searched state being the CETS, hence the ``double quantum'' in the name.
In fact, it is not by chance that the Szegedy operator $W$ is built by means of a 
Grover-like reflection algorithm, which requires the computation of the eigenvalues 
and eigenstates of the Hamiltonian, performed by means of a QPE, as in the QMS.
Moreover, an operation analogous to the one in  Eq.~\eqref{eq:metro_filter_rotation}
is required, for which a single-qubit dedicated register has to be used.

The layout of the quantum state needed by Q$^2$MA is thus the following:
\begin{equation}\label{eq:qqma_state}
\ket{w}_5\ket{\mathrm{acc}}_4\ket{\Delta E}_3\ket{\mathrm{state}}_2\ket{\widetilde{\mathrm{state}}}_1\ ,
\end{equation}
where registers 2 and 1 are used to store the system state and its complex 
conjugate respectively, register 3 is used in the QPE of the energy differences, 
which can be directly computed thanks to the presence of the complex conjugate copy of 
the system, and register 4 is used by an oracle analogous to that in 
Eq.~\eqref{eq:metro_filter_rotation}. 

Once the generalized Szegedy operator 
(obviously depending on $\beta$) is constructed, it can be used as the 
time evolution of a QPE, storing the phases in register 5, on which a classical 
measure is finally performed. If the outcome of this measure is $0$ (which 
corresponds to the eigenvalue 1 of $W$), the input state is projected onto 
the CETS $|\alpha_0(\beta)\rangle$ (up to systematical errors due, e.g., to the 
finite number of qubits adopted in the QPE) and can thus be used to estimate observables. 
If, on the contrary, the measure returns a non-vanishing result, the state
has to be rejected and one should restart the algorithm. 

If the Q$^2$MA procedure is successful, the registers $\ket{w}_5$ and $\ket{\mathrm{acc}}_4$ 
in the final state are both equal to $\ket{0}$, and also $\ket{\Delta E}_3$ is in $\ket{0}$, 
since in the construction of the Szegedy operator $W$, both a QPE in energy and its 
inverse are applied. The entire procedure can be formally thought as the 
application of the projector 
\begin{equation}
\begin{aligned}
\Pi(\beta) =&\ket{\alpha_0(\beta)}\bra{\alpha_0(\beta)}\otimes\ket{0}\bra{0}_5 + \\
&+ \Big[ \sum_{k\neq 0} \ket{\alpha_k(\beta)}\bra{\alpha_k(\beta)} \Big]
\otimes \big[ \Id_5-\ket{0}\bra{0}_5 \big]
\end{aligned}
\end{equation}
to the initial system state, where the $\ket{\alpha_k}$ are the eigenstates of the Szegedy 
operator $W$ (and $\ket{\alpha_0}$ is the CETS), and the subscript 5 refers to 
the ancilla register in which the phase of the $W$-generated QPE is stored.

To increase the probability of measuring the eigenphase $0$ in the QPE using the Szegedy 
operator $W$ (and to reduce the effects of the systematics that will be discussed in 
the following), it is possible to exploit two facts: the first one is that CETS states 
with close temperatures have a good overlap, which differs from one by a quantity of the
order of $(\Delta \beta)^2$~\cite{QQMA_paper}. The second fact is that, in the infinite
temperature limit ($\beta=0$), the CETS is formally equivalent to the maximally
entangled state in the computational basis~\cite{QQMA_paper}, which can be
easily prepared with a combination of Hadamard and C-NOT gates. To obtain with high probability 
the CETS at the desired inverse temperature $\beta$ we
can thus resort to Quantum Simulated Annealing (QSA)~\cite{QSA}, by creating a
sequence of CETS starting from the analytically known one at $\beta = 0$, and then lowering the
temperature in steps of $\Delta\beta = \beta/n_a$, where $n_a$
is the length of the annealing sequence: 
\begin{equation}\label{eq:qsa_sequence}
\ket{\alpha_0^0}\xrightarrow{\Pi_1}\ket{\alpha_0^1}\xrightarrow{\Pi_2}\cdots\xrightarrow{\Pi_{n_a}}\ket{\alpha_0^{n_a}}\ .
\end{equation}
In this equation $\ket{\alpha_0^j}$ is the CETS at $\beta_j=j\beta/n_a$ (with
$j\in\{0,\ldots,n_a\}$) and $\Pi_{j} \equiv \Pi(\beta_j)$.

Therefore, the Q$^2$MA can be summarized as follows:
\begin{enumerate}
\item start at $j=0$ with the maximally entangled state in 
      the computational basis and initialize all ancilla registers 
      to zero;
\item compute the Szegedy operator $W$ corresponding to $\beta_j=j\beta/n_a$, 
      and use it to perform a QPE on the ancilla register 5;
\item perform a classical measurement on $\ket{w}_5$; if the result is 0,
      then proceed to the next step 
      (the state has been correctly projected into the CETS at $\beta_j$), 
      otherwise reset all quantum registers and restart from step 1;
\item iterate steps 2-3 with $j\to j+1$ until $j+1=n_a$.
\end{enumerate}
At the end of the algorithm, we obtain the final state 
\begin{equation}
\ket{0}_5\ket{0}_4\ket{0}_3\sum_i\ \sqrt{e^{-\beta E_i} / Z(\beta)}\ket{\varphi_i}_2\ket{\tilde{\varphi_i}}_1\ ,
\end{equation}
which is equivalent to Eq.~\eqref{eq:rho_th} as far as the probability 
of selecting a state with energy $E_i$ is concerned, i.e., once all the ancilla 
registers 3-5 have been traced out.
In practice, to measure observables on the CETS, one performs a QPE
using the auxilliary register $\ket{\Delta E}_3$, followed by a measure
on the same register, in order to extract with the correct probability energy 
eigenstates on which to perform the measure. Finally, all quantum registers 
are reset and the algorithm is restarted, preparing the CETS for another measurement.

We now come back to the choice of the kick operator $K$. To the best of our knowledge, this 
point has never been fully addressed in the literature, and an operator satisfying all 
the ``optimal'' requirements has always been assumed to exist and to have been selected 
in the discussion of the algorithm. However, the selection of $K$ is 
far from trivial, since such operator has to generate an ergodic selection probability in the 
basis of the Hamiltonian eigenstates, which is obviously unknown for nontrivial problems. 

If $K$ does not generate an ergodic Markov chain, the eigenvalue 1 of the stochastic matrix 
associated with the Markov chain (and thus of the Szegedy operator $W$) can have 
larger than one degeneracy. This implies that the projection on the phase zero of 
the Szegedy-generated QPE does not ensure the selection of the CETS.
An operative way to eliminate this problem, or at least to reduce 
its consequences, is to use different kick operators in the different 
annealing steps $j=1,\ldots, n_a$, thus projecting at step $j$ on the eigenspace 
corresponding to the eigenvalue 1 of the Szegedy operator $W_j$ built using the 
kick operator $K_j$. Since the CETS always corresponds to an eigenstate with eigenvalue 1 
of the Szegedy operator for any $K_j$, in this way we expect to 
correct the ergodicity problem of the single-kick \emph{naive} implementation of the Q$^2$MA.

Let us stress that, in principle, one should require all the kick operators, and 
the corresponding Szegedy operators, to be considered at the same time for each 
annealing step, in order to guarantee that CETS is the only possible outcome in the case 
of acceptance. However, the implementation in which different single kick operators 
are used in different annealing steps is expected to work well in practice, at least 
as long as there is a large overlap between CETS corresponding to different annealing steps,
i.e., as long as the annealing procedure is slow enough. In particular it is reasonable to 
guess that the convergence of the annealing process scales as $1/n_a$, provided that $n_a$ is larger 
than the minimum number of Szegedy operators required to univocally identify the CETS. Since 
the aim of the algorithm is to generate a (large) sample of measures from which to extract 
averages and standard errors, the operators $K_j$ could be also generated stochastically, as 
long as the CETS is identified almost surely (i.e., with probability 1).

\subsection{Sources of systematical errors}
\label{subsec:expected_syst}

Several sources of systematical errors exist both in the QMS and in the Q$^2$MA 
simulation schemes. In the following, we try to understand their impact on the
simulation results in a controlled setting, by using a simple toy model
to investigate separately the different contributions.

To avoid over-complicating our analysis, and since here the focus is on
the quantum algorithms themselves, we are going to consider a basic
spin model, for which no digitization error is present, 
unlike more complex systems, like QCD and systems with continuous gauge symmetries.

It should be clear that a common source of systematical error, 
in both QMS and Q$^2$MA, is also the digitization error of the energies 
used in the QPE performed by using the system Hamiltonian. In the 
QMS this step is required by the oracle of Eq.~\eqref{eq:metro_filter_rotation}, 
while in the Q$^2$MA this step is hidden in the construction of the generalized 
Szegedy operator~\cite{QQMA_paper}. Note that problems related to the accuracy 
of the energy representation adopted are present virtually in any importance 
sampling Monte Carlo computation (also classical ones), although this issue is 
usually overlooked in most practical computations \cite{Roberts:1998, Breyer:2001}. 
Since our aim is to test the effectiveness of quantum algorithms in computing thermal averages, 
the sensitivity to the energy digitization is a relevant property to be investigated.
For the model we study (as for all two-level systems), it is however possible to 
find an encoding in which energies have no digitization error. This is 
obviously impossible for generic systems with incommensurate energy levels, 
but it allows us to focus on other sources of systematical error
which are instead intrinsic of the algorithms we are studying 
(i.e., not simply inherited by the energy QPE). 
In the following, when not specified, we will always assume 
that such an exact encoding of the energy is being used.
We will investigate what happens when such a constraint is relaxed 
in a later section.

The systematical error that is specific to the QMS, and which is avoided
by the Q$^2$MA, is the one related to the thermalization
time of the algorithm: Hamiltonian eigenstates are sampled with the Boltzmann
statistics only asymptotically for large times, i.e., after many iterations. 
The leading correction to this asymptotic behavior scales as $\exp[-(1-\lambda_2)t]$, 
where $t$ is the number of Monte Carlo steps (the so called the Monte Carlo time)
and $\lambda_2$ is the second largest eigenvalue of the Markov-chain transition matrix,  
the largest eigenvalue being 1 by construction. 
In a classical Monte Carlo sampling, nothing but the computational power prevents 
us from using very long Markov chains, and the algorithm is thus stochastically exact, since 
typical statistical errors scale as $1/\sqrt{t}$. 
This is the case also for the QMS only if one is interested in the thermal
averages of observables compatible with the Hamiltonian. 
If instead the aim is to compute $\langle A\rangle$, with an operator $A$ which does not commute 
with the Hamiltonian, the Markov chain breaks down: by measuring $A$ on 
the energy eigenstate $\ket{\varphi_k}$ 
extracted by the QMS, the state is projected to an eigenvector of $A$, which is generically
a linear combination of different energy eigenstates.
Therefore, after such a measure, the QMS needs to be reinitialized. 
The number of updates performed between subsequent measures of $A$ 
thus puts an upper bound on the attainable accuracy.

In the Q$^2$MA, QPE is used not only with the time evolution 
generated by the Hamiltonian, but also with the unitary Szegedy operator. 
In this case, the typical systematical error is the digitization error 
on the phase of this evolution, related to the number of qubits
used for the register $\ket{w}_5$ in Eq.~\eqref{eq:qqma_state}.
We noted before that, for simple enough systems, it is possible to find
an exact encoding of the energies, thus removing the systematic associated 
to the Hamiltonian related QPE. An analogous procedure is generically not possible 
for the QPE with the Szegedy operator $W$ as time evolution, since $W$ and its eigenvalues 
depend on the (inverse) temperature value $\beta$. 

Analogously as for the QMS, which becomes exact in the large-time limit, the Q$^2$MA is also exact in the adiabatic limit of infinite annealing steps, provided that the 
single kick $K$ generates an ergodic chain. Indeed, if we denote by $\epsilon$ the 
magnitude of the error introduced on the state by the inaccuracy of the Szegedy QPE, 
the global error at each step of the annealing procedure is of the order of 
$\epsilon(\Delta\beta)^2=\epsilon\beta^2/n_a^2$ 
and the final error is thus $O(\epsilon\beta^2/n_a)$~\cite{QQMA_paper}. Note that if the Szegedy 
QPEs were performed exactly (i.e., $\epsilon=0$), the algorithm would be exact for any number 
of annealing steps $n_a$; however, the success probability of step 3 in Sec.~\ref{subsec:qqma} 
would be extremely small if $n_a$ is not large enough.

When different kick operators $K_i$ are used to effectively restore ergodicity (see the 
discussion at the end of Sec.~\ref{subsec:qqma}) this is a priori no more true: the eigenspace 
corresponding to the eigenvalue 1 of each Szegedy operator $W_i$ has dimensionality larger than one, 
so the global error at each step of the annealing procedure can be, 
at least in principle, of the order of 
$\epsilon \Delta\beta=\epsilon\beta/n_a$. That would make the final error independent of 
the number of annealing steps, i.e., the quantum annealing would just increase the success 
probability of the CETS generation, with no effect on the systematical
error, so that the only way of removing systematics from the final estimate would 
be to reduce $\epsilon$, i.e. to reduce the inaccuracy of the Szegedy QPE.

Contrary to such expectations, as we will show in the following, the systematic 
errors of the Q$^2$MA algorithm appear instead to depend on the number of annealing steps, 
as if a single and ergodic kick operator were used. A possible interpretation is that, 
once the system collapses onto the correct CETS during the annealing procedure 
(actually one surely starts from the correct CETS at $\beta = 0$), it is highly 
probable to keep staying on CETS, due to the good overlap of CETS states at different 
annealing steps, thus making single step errors of $O(\Delta \beta^2)$ again.

\section{Model and metrics}
\label{sec:model}

The first part of this section is dedicated to a description of the 
particular quantum system used as a test bed for our analysis. Then we 
introduce some quantities that will be used to quantify the systematical 
errors of the explored quantum algorithms.

\subsection{The frustrated triangle}\label{subsec:frutri}

To compare the QMS and Q$^2$MA 
algorithms described in the previous sections, we consider a 
system of three quantum spin-1/2 variables with Hamiltonian~\cite{Qubipf_QMS}
\begin{equation}
  H = J ( \sigma_x \otimes \sigma_x \otimes \Id + \sigma_x 
  \otimes \Id \otimes \sigma_x  +  \Id \otimes \sigma_x \otimes \sigma_x )\, ,
  \label{eq:HamSpinSys}
\end{equation}
where $\sigma_j$ stands for the usual Pauli matrices,
$\Id$ is the $2 \times 2$ identity operator and the coupling $J$ 
is positive (i.e., antiferromagnetic), to make the system frustrated.

It is easy to find a basis of the Hilbert space which makes the problem 
trivial: when going on the basis in which all $\sigma_x$ operators are diagonal, 
it is immediate to see that two distinct degenerate energy levels exist: the 
fundamental one with energy $E_0 = -J$ and degeneracy 6 (corresponding to 
the case of two spins aligned) and the excited one with energy $E_1 = 3 J$ and degeneracy 2 
(corresponding to three spins aligned). 
However, to mimic a realistic situation, we work in the standard 
computational basis, where all $\sigma_z$ operators are diagonal. 
Using this basis to evaluate the thermodynamical quantities by means 
of the Trotter--Suzuki decomposition, it is possible to verify 
that here the standard path-integral 
importance sampling Monte Carlo fails,
due to a sign problem (see Ref.~\cite{Qubipf_QMS}). This problem is
obviously absent in the quantum computational approach.

As discussed in Secs.~\ref{subsec:qms}-\ref{subsec:qqma}, both QMS and Q$^2$MA 
require the application of unitary operators to sample the state space: in the QMS, we 
need the set of operators denoted by $\mathcal{C}$ in Sec.~\ref{subsec:qms} to evolve 
the Markov chain; for the Q$^2$MA, we need the kick operators $K_i$ to build the Szegedy 
operators $W_j$ (see Sec.~\ref{subsec:qqma}). For both algorithms, 
the adopted unitary operators are Hadamard gates ($\Had$) acting on a single qubit of the register of the system state, i.e., $\mathcal{C} \equiv \{K_0, K_1, K_2\}$,
where $K_0 = \Id \otimes \Id \otimes \Had$, 
$K_1 = \Id \otimes \Had \otimes \Id$, and $K_2 = \Had \otimes \Id \otimes \Id$.

\subsection{Quantifying systematical errors}\label{subsec:systerr}

Systematical errors induce biases in the thermal averages computed by means of 
QMS or Q$^2$MA. In the simple test system adopted here these biases can be identified 
by comparing the numerically estimated values with the analytically-known exact 
ones. Since systematic errors affect in a different way the various observables, 
to quantify them we decided to use the following three metrics:
\begin{enumerate}
\item 
the bias in the expectation value of the Hamiltonian (i.e., in the internal energy):
\begin{equation}\label{eq:discr_Ene}
    d_{\text{Ene}}\equiv \lvert \bar{E} -{\langle E \rangle}_{\text{exact}} \rvert \ ,
\end{equation}
where $\bar{E}$ is the mean energy estimated by using the quantum 
algorithm, while $\langle E\rangle_{\text{exact}}$ is the exact 
expectation value of the energy at (inverse) temperature $\beta$, given by 
\begin{equation}
\langle E\rangle_{\text{exact}}=3J\frac{e^{-3\beta J}-e^{\beta J}}{e^{-3\beta J}+3e^{\beta J}}\ ;
\end{equation}

\item
the bias in the expectation value of an observable $A$ which does not commute 
with $H$. For this purpose, we choose 
$A \equiv \sigma_x \otimes \sigma_x \otimes {\big( \Id+ \sigma_y\big)}$ 
and define
\begin{equation}\label{eq:discr_Aop}
    d_{\text{Aop}}\equiv \lvert \bar{A} -{\langle A \rangle}_{\text{exact}} \rvert \ ,
\end{equation}
where $\bar{A}$ is the value estimated by using the quantum 
algorithm and $\langle A\rangle_{\text{exact}} = 
\langle E \rangle_{\text{exact}} /3$ is the analytically known result;

\item
the distance in the space of the density matrices
\begin{equation}\label{eq:discr_TrD}
    d_{\text{TrD}}\equiv \frac{1}{2} \|\bar{\rho} - \rho_{\text{exact}}\|_1 \ ,
\end{equation}
where for a matrix $M$ the norm $\|M\|_1=\mathrm{Tr}\sqrt{M^{\dag}M}$ is used. 
\end{enumerate}

The figure of merit in Eq.~\eqref{eq:discr_TrD} is particularly significant,
since it directly quantifies the bias in the probability distribution 
and not only in some specific average value, which could be small by chance. 
Its definition requires however some comments.
While the computation of $\bar{E}$ (and analogously of $\bar{A}$) can be 
carried out (at least from a theoretical point of view) on a quantum simulator
by using $\bar{E}=\frac{1}{N}\sum_{i=1}^N E_i$, where $E_i$ is the energy 
observed in the $i$-th draw, this is not the case for $\bar{\rho}$, 
since it is not possible to measure at once 
the density matrix $\rho_i$ of the $i$-th draw.
However, since we are testing the algorithms using a quantum simulator
and not a real quantum machine, for the purpose of investigating the systematical errors 
we can actually pick up the full state of the algorithm, extract $\rho_i$, and compute 
$\bar{\rho}\equiv \frac{1}{N} \sum_i \rho_i$.

\section{Results}\label{sec:numres}

The particular features of the explored model permits to 
disentangle effects related to the inexact energy 
representation from the QPE from other systematics. 
For this reason, in the first part of this section we work
within an exact energy representation scheme, switching then to the inexact one in the second part.

The practical implementation of the explored algorithms is based on a quantum 
emulator running on GPUs (Simulator for Universal Quantum Algorithms, SUQA) 
developed by one of the authors (GC), which is inspired to the well known Qiskit simulator.

\subsection{Exact energy representation}

The frustrated triangle can be described with an exact encoding 
of its degrees of freedom using three qubits only. 
Moreover, as mentioned in Sec.~\ref{subsec:frutri}, the system has 
two energy levels, which can be represented exactly with one qubit 
for the energy registers, thus removing any systematical error 
related to the QPE with the Hamiltonian. 
Energy differences (needed for the Q$^2$MA) can thus be exactly represented by 
using two qubits. 

As previously discussed, when this exact energy encoding
is used, QMS and Q$^2$MA have different sources of systematical errors, 
that will be investigated below.

\subsubsection{Quantum Metropolis Sampling}\label{subsubsec:systerr_qms}

As discussed in Sec.~\ref{subsec:expected_syst}, the QMS is not stochastically exact when used to 
compute the thermal average of an observable which does not commute with the Hamiltonian
(as for the operator $A$ defined in Sec.~\ref{subsec:systerr}), 
since the measurement of $A$ breaks the Markov chain evolution.
The only parameter which controls the size of the systematical error 
introduced by this breaking is the number of updates performed 
between consecutive measures of $A$. 

\begin{figure}[!t]
  \includegraphics[width=1.\linewidth, clip]{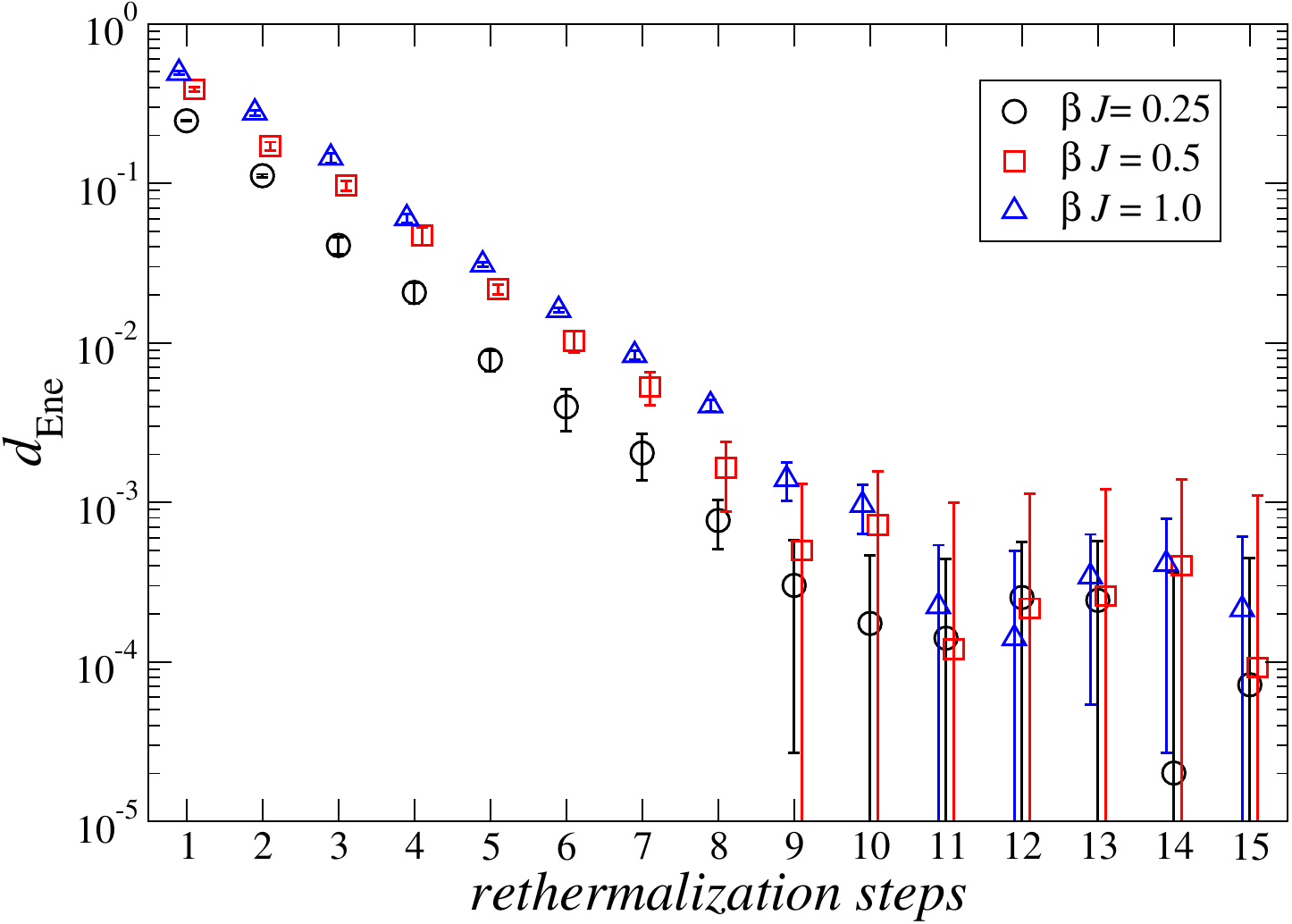} \vspace*{1mm}
  \includegraphics[width=1.\linewidth, clip]{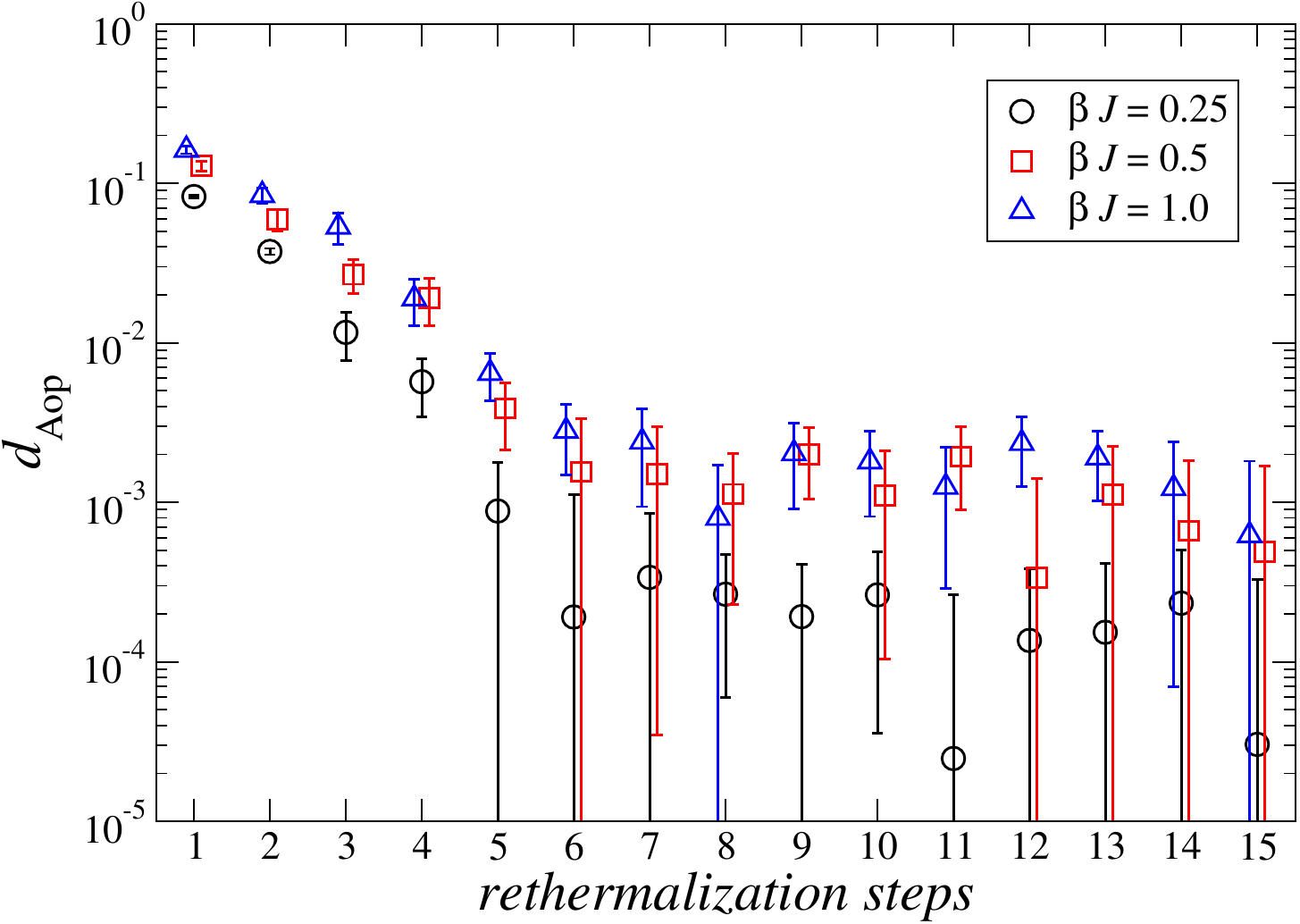}\vspace*{1.5mm}
  \includegraphics[width=1.\linewidth, clip]{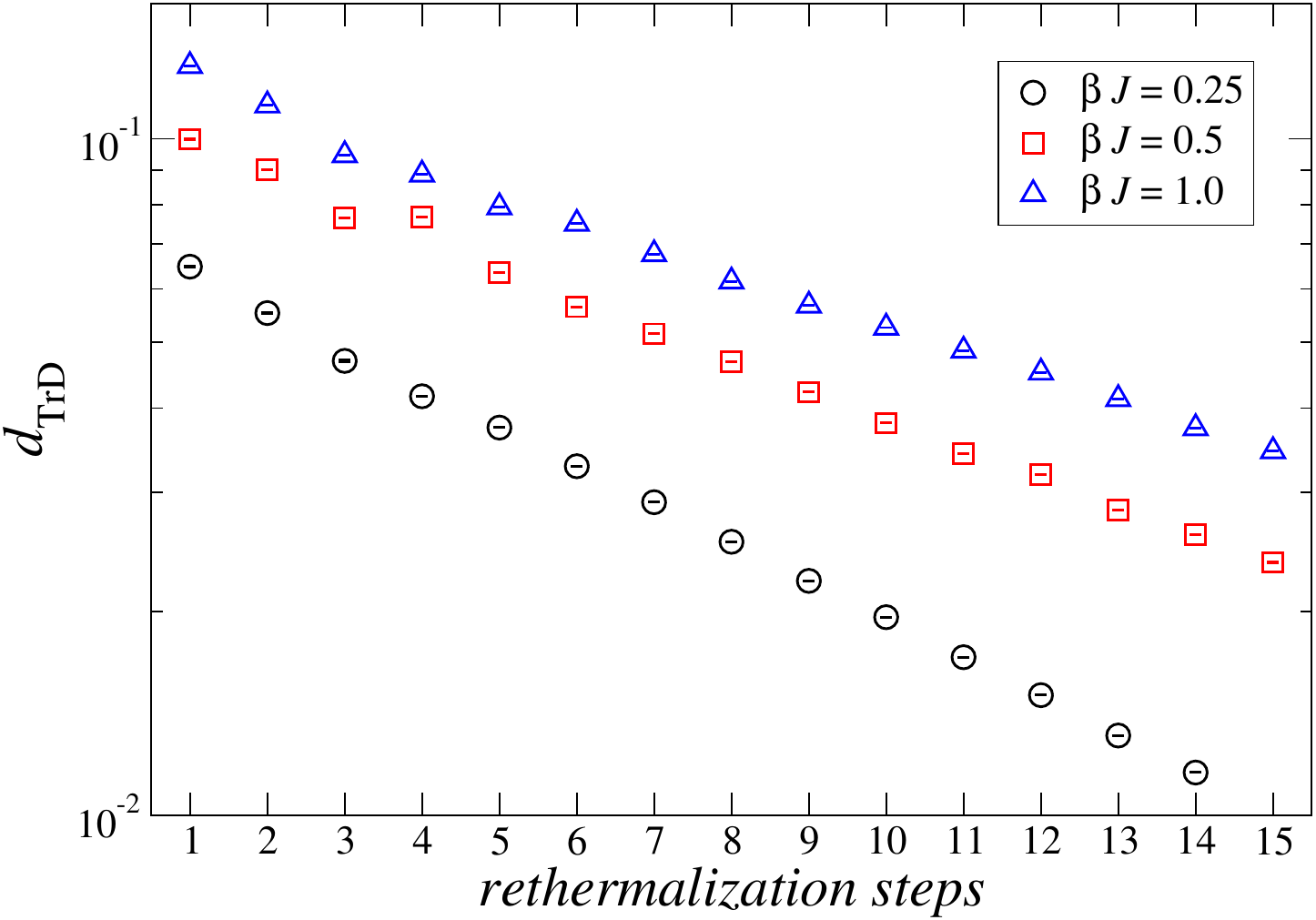}
  \caption{QMS: behavior of $d_{\text{Ene}}$ (top panel),
      $d_{\text{Aop}}$ (middle panel),
      and $d_{\text{TrD}}$ (lower panel),
      as a function of the number of rethermalization steps,
      for $\beta J= 0.25, 0.5$, and $1.0$.}
    \label{fig:systest_qms}
\end{figure}

Once a measurement of $A$ has been performed, two different strategies are
possible: one can either restart the chain from the beginning
or make the state (which is now an eigenstate of $A$) collapse back to 
an energy eigenstate (possibly different from the original one), by performing a QPE 
on the energy register followed by a measurement~\cite{Qubipf_QMS}. 
In the first case, the number of updates between different measurements 
is called \emph{thermalization time}, while in the second case 
it is more natural to call it \emph{rethermalization time}. 
As in Ref.~\cite{Qubipf_QMS}, we follow the second strategy,
which seems to be slightly more efficient.

The density matrix after the $i$-th rethermalization step is evaluated, before 
breaking the state with the $A$-measurement, by reading the state from the system 
register $\ket{\psi_{k(i)}}$ and building the projector 
$\rho_i \equiv \ket{\psi_{k(i)}}\bra{\psi_{k(i)}}$. 
Of course, any single instance of $\rho_i$ will be far from the exact density 
matrix $\rho_{\text{exact}}$, irrespective of the number of rethermalization steps. 
However, in the large sample limit, the (non-vanishing) discrepancy between $\bar{\rho}$ and $\rho_{\text{exact}}$ is only due to the systematic introduced by rethermalization
and it is expected to vanish in the limit of an infinite number of rethermalization steps.

Results obtained for the three accuracy metrics $d_{\text{Ene}}$, $d_{\text{Aop}}$, 
and $d_{\text{TrD}}$ introduced in the previous section are shown in Fig.~\ref{fig:systest_qms}
(respectively top, middle, and bottom panel), for three values of the inverse 
temperature $\beta J=0.25, 0.5$, and $1.0$. The statistics accumulated for the 
different points is not homogeneous, since runs have been stopped when $d_{\text{TrD}}$ 
reached a fixed accuracy. For this reason, runs performed at larger values of 
rethermalization steps, for which the bias is smaller, are significantly longer than 
the ones performed at smaller values. The stopping criterium adopted, together with 
the fact that $d_{\text{TrD}}$ is the most observable-independent among the adopted 
metrics, also explains why data reported in the top and central panel, respectively 
for $d_{\text{Ene}}$ and $d_{\text{Aop}}$, have significantly larger relative errors 
compared to those in the bottom panel, for $d_{\text{TrD}}$.
It is also clear that, for all the metrics considered, the systematic bias approaches 
zero exponentially in the number of rethermalization steps, as expected on theoretical grounds.

\subsubsection{Quantum Quantum Metropolis Algorithm}\label{subsubsec:systerr_qqma}

\begin{figure}[!t]
\includegraphics[width=1.0\linewidth, clip]{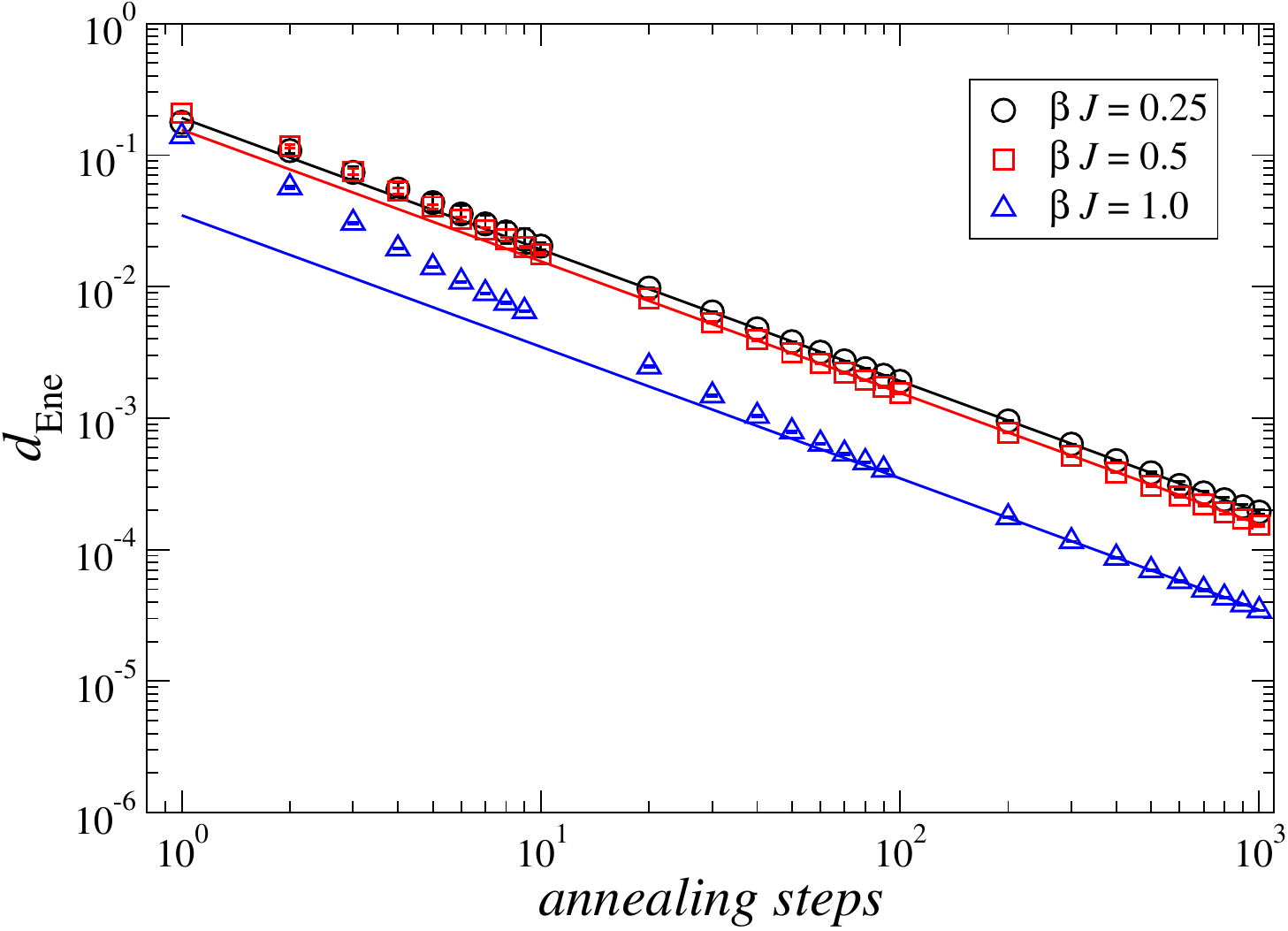}
\includegraphics[width=1.0\linewidth, clip]{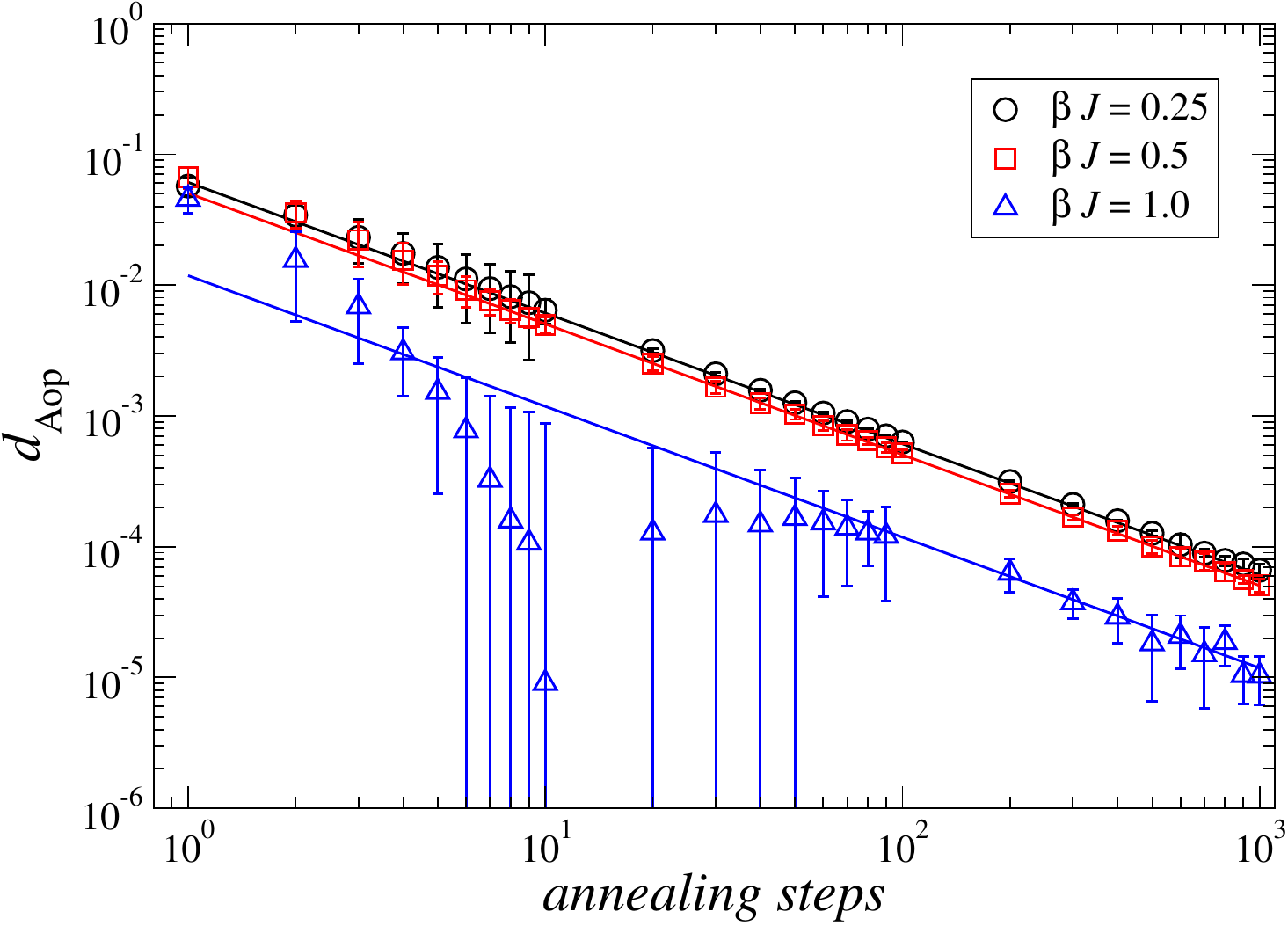}
\includegraphics[width=1.0\linewidth, clip]{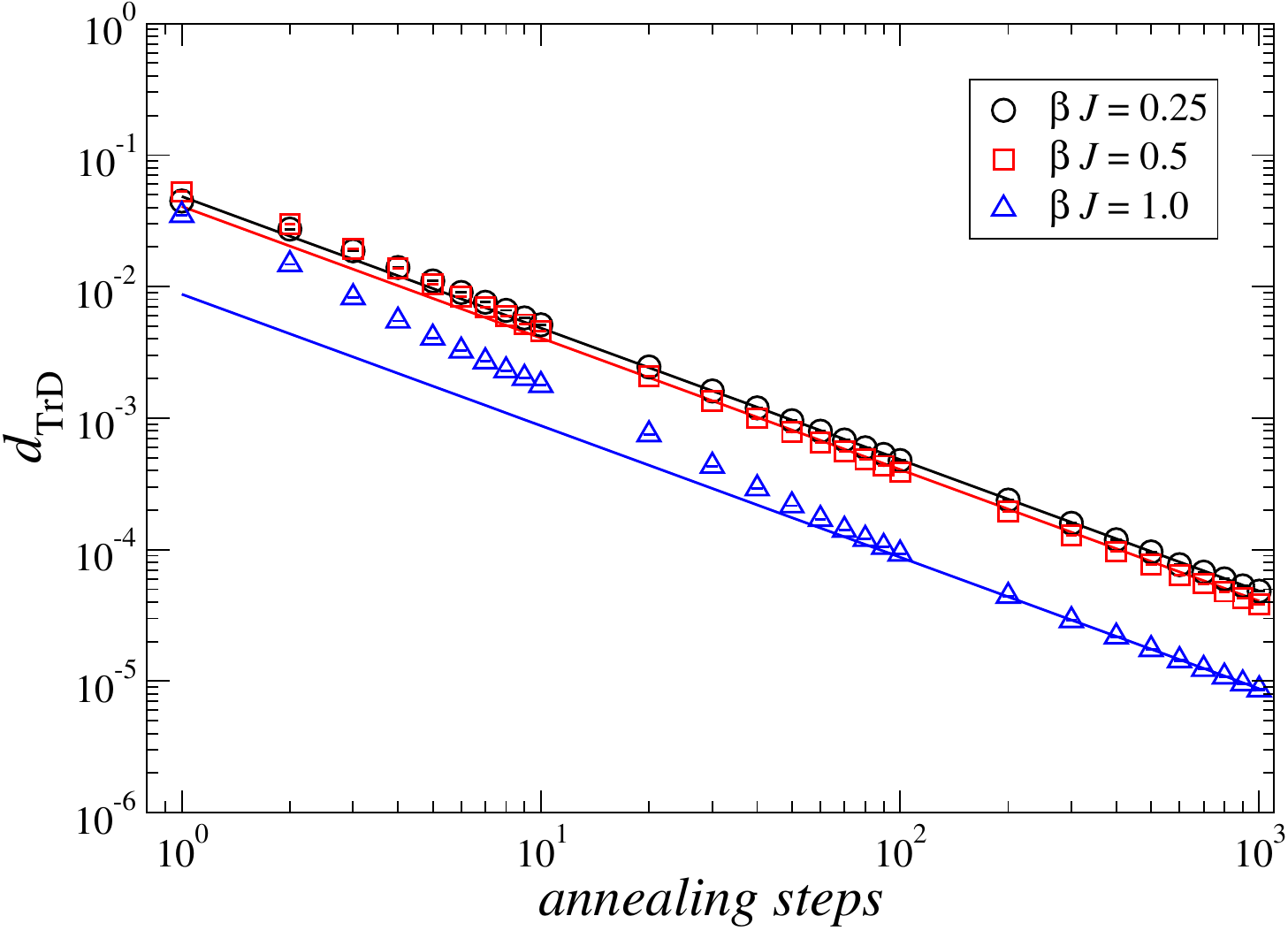}
\caption{Q$^2$MA: behavior of $d_{\text{Ene}}$ (top panel),
$d_{\text{Aop}}$ (middle panel),
and $d_{\text{TrD}}$ (bottom panel),
 as a function of the annealing steps $n_a$, for $\beta J = 0.25, 0.5$ and $1.0$,
when using 3 qubits in the register $\ket{w}_5$, see Eq.~\eqref{eq:qqma_state}.
Continuous lines represent fits to the $1/n_a$ behavior, which well 
reproduces data at large enough $n_a$.}
\label{fig:systest_qsa}
\end{figure}

As explained in Sec.~\ref{subsec:expected_syst}, when using an exact energy 
representation, the systematic errors of the Q$^2$MA algorithm are only due to 
the inexactness of the Szegedy QPE and to the finite number of annealing steps. 
However, we recall that it is generally not possible to make the Szegedy QPE 
exact, as the $W$ operator depends on $\beta$. 
Results presented in this section have been obtained by using the range $[0.5, 1]$ for the 
Szegedy QPE (we are only interested to the eigenvalue 1) and fixing the number of qubits in the $\ket{w}_5$
register to 3, studying the dependence of systematic errors on the number 
of annealing steps. Consistent results are obtained when using 4 qubits for the register
$\ket{w}_5$.

Contrary to what happens in the QMS case, it is not possible to measure $A$ and 
$E$ during the same run, and a new CETS reconstruction is required after each 
measurement. Since here we are only interested in systematics related to the 
incorrect determination of CETS, we evaluated 
the density matrix after each run of the algorithm, then exploiting the  use of
an emulator (rather than a real machine) to determine the exact average 
values of $A$ and $E$ corresponding to the given density matrix. Therefore, 
statistical errors shown in the following analysis are only due to fluctuations 
in the CETS determination from run to run, and are in fact small and well below 
the symbol size in most cases.

In Fig.~\ref{fig:systest_qsa}, we present the results for $d_\text{Ene}$ (top panel), 
$d_\text{Aop}$ (middle panel), and $d_\text{TrD}$ (bottom panel) for the same three values of 
temperature explored in the previous subsection. In order to achieve a target precision on 
$d_\text{TrD}$, the runs performed with higher numbers of annealing steps required more 
Q$^2$MA iterations, as the systematic error decreases with increasing steps.
It can be observed that error bars are visible and not homogeneous only in the middle panel, 
where the systematic error also shows a non-monotonic behavior as a function of the number 
of annealing steps $n_a$: this is partially due to a change of sign in the bias of $A$ as 
a function of $n_a$, which occurs accidentally for this particular choice of parameters.

A clear difference of Fig.~\ref{fig:systest_qsa}, with respect to Fig.~\ref{fig:systest_qms}, 
is that the scale is logarithmic on both axes. The reason is that in this case systematics 
appear to decrease polynomially (instead of exponentially) with the number of annealing steps. 
In particular, it is interesting to notice that, for large enough $n_a$, i.e., when the annealing 
step is small enough, data are well compatible with a $1/n_a$ behaviour.
This result agrees with the prediction reported in Ref.~\cite{QQMA_paper} for a single and 
ergodic kick operator. However, in the light of the discussion reported in Sec.~\ref{sec:algo}, the 
result is far from trivial, and can be interpreted heuristically as evidence that, randomly 
alternating different kick operators in the different 
annealing steps, effectively reproduces, in the large annealing step limit, the ideal 
behaviour predicted for a single ergodic kick operator.

\subsection{Inexact energy representation}\label{subsubsec:systerr_inexactenergy}

\begin{figure}[!t]
\includegraphics[width=1.0\linewidth, clip]{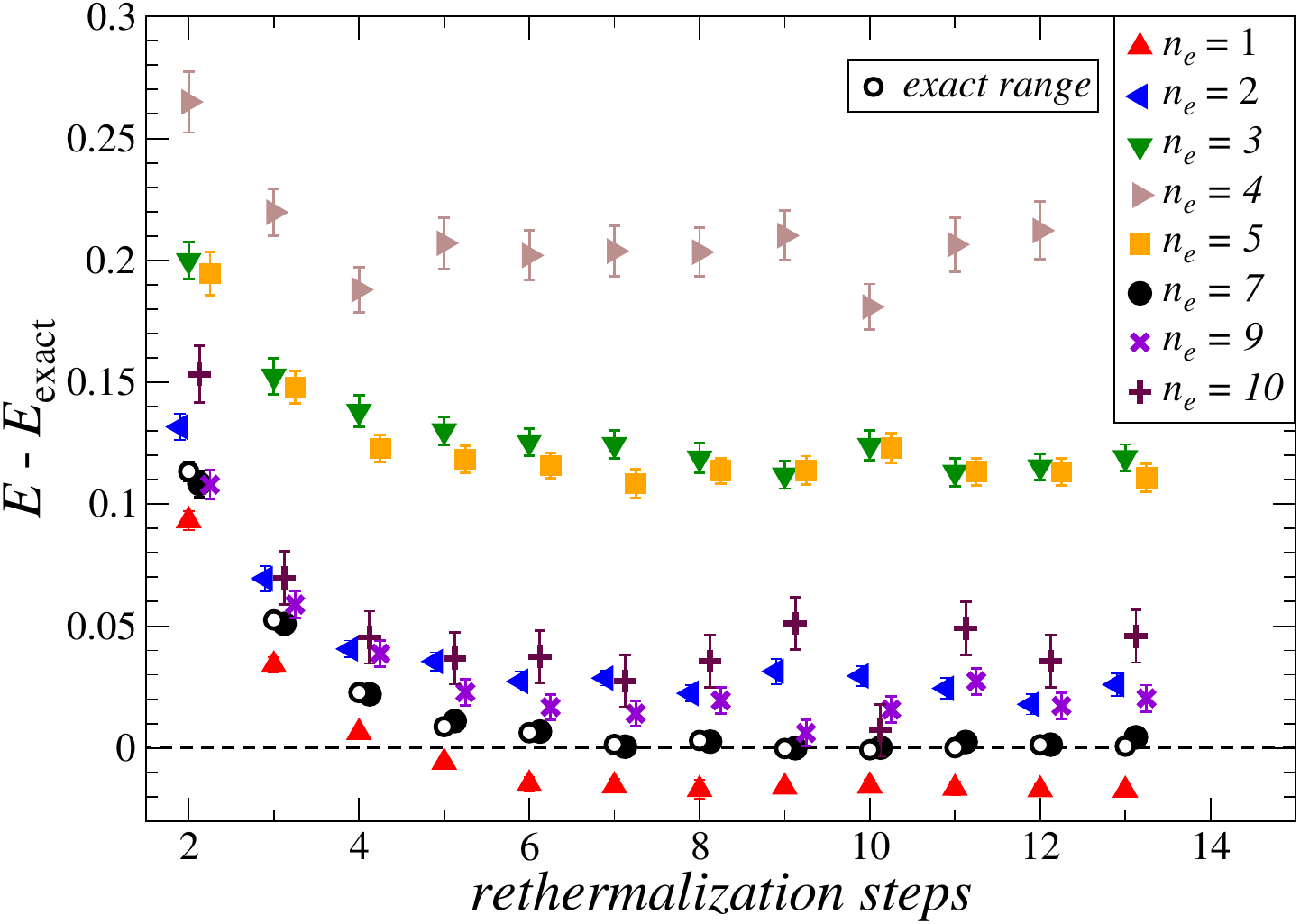}
\caption{Energy discrepancy for the QMS algorithm with inexact QPE grid with fixed extrema $[-1.1,3.1]$ 
(in units of $J$) at $\beta J= 0.25$.}
\label{fig:qms_wrongrange_systematic}
\end{figure}

In the previous analysis, working with an exact energy representation was instrumental
to isolate some algorithmic-specific sources of systematical errors. This was possible 
because the considered quantum system has only two different exactly known energy levels, 
but it is clearly unrealistic in any case of direct physical interest. 
In such general cases, one has to use an inexact energy representation with $n_e$ qubit in the energy register(s), 
thus introducing further systemtatics.

Denoting by $E_0$ and $E_1$ the exact energy eigenstates of the 
frustrated triangle system (see Sec.~\ref{subsec:frutri}), we consider for the QPE an 
interval larger then $[E_0, E_1]$ by an amount of about $2\delta$. Two natural 
prescriptions exist to place the $2^{n_e}$ grid points of the QPE: the ``fixed extrema 
grid'' and the ``refined grid'' cases.

In the ``fixed extrema'' case the inteval is fixed to be $[E_0-\delta, E_1+\delta]$ and the $2^{n_e}$ points
are uniformly distributed in this interval. As $n_e$ is increased, the grid becomes finer, but the position 
of all the grid-points changes with $n_e$. On the other hand, in the ``refined grid'' case we use the interval
\begin{equation}
[E_0-\delta,E_1+\delta+(E_1-E_0+2\delta)(1-2^{1-n_e})]\ ,
\end{equation}
which is chosen in such a way that, by increasing $n_e$, all points of the coarser grid 
are also present in the finer grid (with exception of the largest value).

Figure~\ref{fig:qms_wrongrange_systematic} shows the behavior of $\braket{E}$ as a function of the number of 
rethermalization steps for different values of $n_e$ in the ``fixed extrema" scheme. 
It is clear that, for a large enough number of retermalization steps, a plateau emerges in $\braket{E}$,
signaling the presence of a systematical difference between $\braket{E}$ and $\braket{E}_{\mathrm{exact}}$.
This systematic is expected to vanish for $n_e\to\infty$, however the approach to the large $n_e$ limit is 
obviously non monotonic, at least in the range of $n_e$ values explored. This could make the diagnostic of 
the convergence non trivial in realistic cases, in which the exact expectation value is not known and the number 
of values of $n_e$ is limited by the hardware capability.

\begin{figure}[!t]
\includegraphics[width=1.0\linewidth, clip]{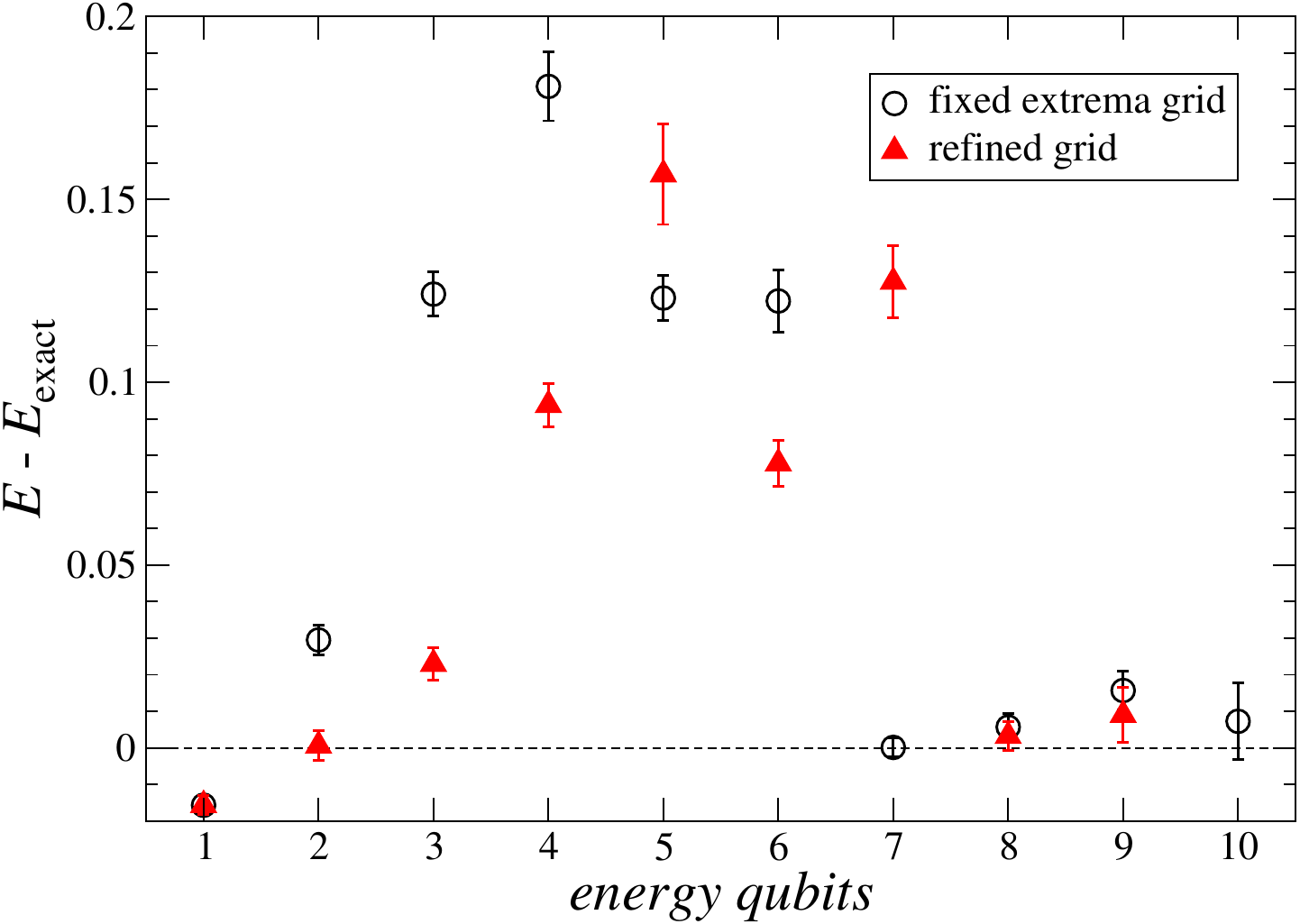}
\caption{Effects of the two grid prescriptions for the QMS algorithm on energy discrepancies with 
inexact QPE grids at $\beta J= 0.25$ and with 10 rethermalization steps.}
\label{fig:qms_fixrange_vs_refinegrid}
\end{figure}

For this reason we also investigated the ``refined grid'' method, which could be expected to have a smoother 
approach to the large $n_e$ limit. This is however not the case, as can be seen from 
Fig.~\ref{fig:qms_fixrange_vs_refinegrid}, in which a comparison of two approaches is performed using 
10 rethermalization steps (note that this number of retermalization steps is well in the plateau of Fig.~\ref{fig:qms_wrongrange_systematic}). Both the approaches show a non monotonic scaling for an intermediate range 
of $n_e$ values, and converge to the correct asymptotic result for $n_e>8$. It is however reasonable to guess the 
non monotonic behaviour to continue also for larger values of $n_e$, where it is hidden by the statistical 
accuracy of our data.

\begin{figure}[!t]
\includegraphics[width=1.0\linewidth, clip]{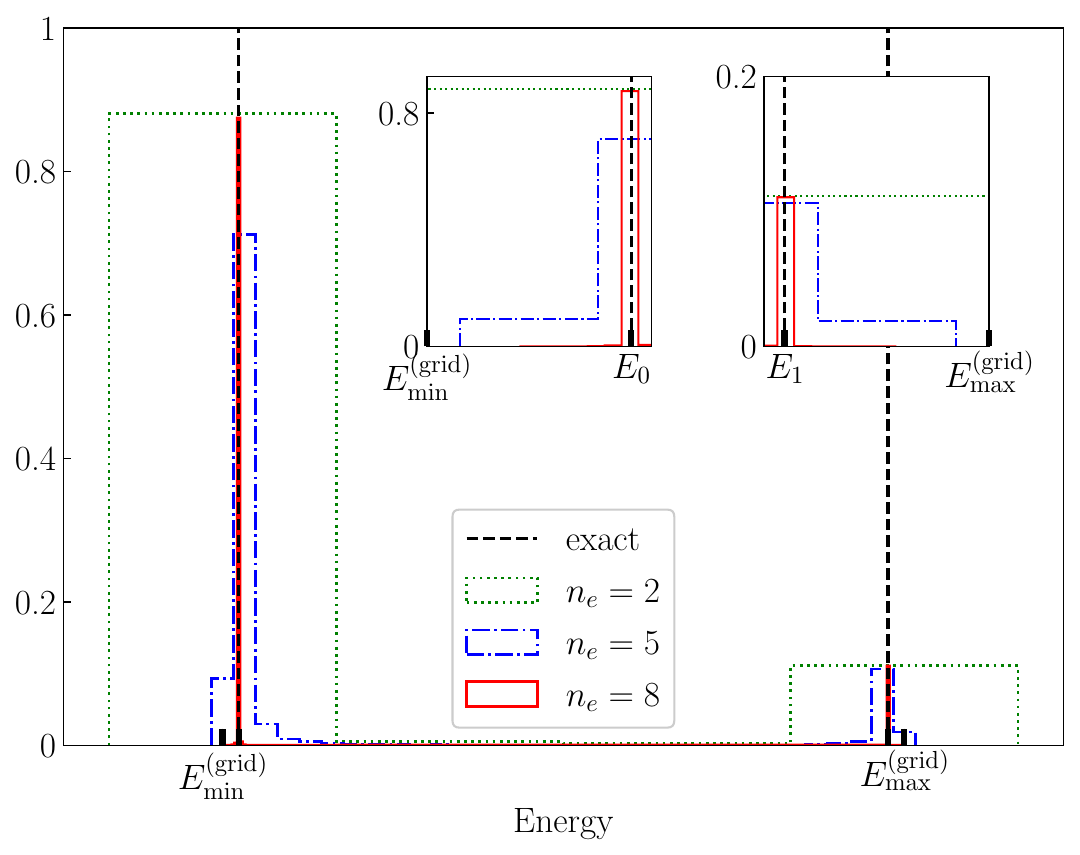}
\caption{Comparison between the exact energy distribution and the one computed from the QMS 
using an energy grid in the QPE with fixed extrema and with inexact range 
$[E_{\min}^{(\mathrm{grid})}, E_{\max}^{(\mathrm{grid})}] = [-1.1,3.1]$ (in units of $J$), 
for 3 values of the number of qubits of the energy register $n_e$. Figure refers in all cases 
to measures separated by 10 rethermalization steps. Insets provide a zoom over the regions 
close to the exact energy levels $E_0$ and $E_1$.}
\label{fig:qms_histo_rt13}
\end{figure}

The reason for the non monotonic behaviour is related to the fact that, for a given value of $n_e$,
an eigenvalue can by chance be well-inside one of the QPE grid-interval
or close to the boundary between two consecutive grid-intervals. Which of the two cases happen depends on $n_e$ and
changes by increasing $n_e$, as can be seen 
in Fig.~\ref{fig:qms_histo_rt13}. Obviously the oscillations induced by this effect gets smaller and smaller as
$n_e$ is increased, and ultimately convergence is reached with an arbitrary accuracy, but the approach to the 
asymptotic value presents oscillations. This effect is particularly evident in the system studied in this paper 
since the spectrum is very simple, consisting just of two points. In more complex systems, with a less simple 
spectrum, it seems reasonable to assume this discretization effect to be less 
significant, with some form of self-averaging happening. 

We tried to repeat a similar analysis to investigate the effect of an inexact energy representation 
also in the Q$^2$MA case, however the results turned out to be much less clear, a fact that is probably 
due to several aspects. First of all the way in which energies enter the construction of the 
Szegedy operator is much more involved that the way in which they enter the Metropolis filter in the QMS 
algorithm, so ``error propagation'' is nontrivial for the limited number of qubits that can be used in the simulator. 
Another important point is the fact that in the Q$^2$MA also other sources of systematics are present, like the 
discretization error in the Szegedy-related QPE, and different systematics can interact in a nontrivial way with each other. 
For this reason we were not able to identify a reasonable trend in our data for the case of the inexact energy representation
in the Q$^2$MA.

\section{Discussion and conclusions}\label{sec:conclusions}

This study is a step along a research line dedicated to the exploration 
of quantum algorithms for the computation of quantum thermal averages,
in view of future applications to complex and interesting physical systems, like the 
fundamental theory of strong interactions, when technological developments will allow 
for reliable and scalable quantum machines.

In Ref.~\cite{Qubipf_QMS}, some of the present authors already explored the Quantum Metropolis Sampling (QMS) algorithm and applied it to a frustrated three-spin system. In this case, we have considered the same physical system to develop a practical implementation of the Quantum-Quantum Metropolis Algorithm (Q$^2$MA), proposed in Ref.~\cite{QQMA_paper}. This algorithm in principle represents a conceptual improvement over the QMS, as it enjoys a further quantum advantage. Such advantage stems from the fact that, while the QMS performs a Markov chain among the quantum states of the system which is classical in spirit and which faces the difficulties
of the no-cloning theorem, the Q$^2$MA acts like a quantum searching algorithm, where the searched
state is the so-called CETS, which is a pure state in a doubled Hilbert space, whose amplitudes encode the thermal distribution of the target system. 
Provided that the CETS is found, the algorithm is exact in principle;
however, an annealing procedure, in which one finds iteratively different CETS states corresponding 
to a descending sequence of temperatures, is used to improve the success probability of the searching algorithm.

A first result of our investigation is that the practical implementation of the 
Q$^2$MA algorithm might be far less trivial. The searching algorithm is based on the construction 
of a Szegedy operator, which is assumed to have a single eigenstate with 
eigenvalue $\lambda = 1$, corresponding to the CETS, which is actually the 
eigenstate the algorithm looks for. On the other hand, the construction of the Szegedy
operator is based on the definition of a kick operator $K$, which is representative
of a Markov chain and should be ergodic to guarantee the non-degeneracy of the 
$\lambda = 1$ eigenstate: if this is not the case, the algorithm is not guaranteed to 
find the correct CETS, leading to possible systematics.

As we have discussed in Sec.~\ref{sec:algo}, finding an ergodic kick operator is a non-trivial assumption.
However, a possible conceptual modification of the algorithm is to make use of different
kick operators, which are randomly alternated during the annealing procedure:
even if the single kick operators are non-ergodic, the fact that CETS corresponding 
to close enough temperatures have a good overlap is expected to strongly enhance
the probability that the correct CETS is found along the annealing sequence,
at least if the annealing step is small enough.

As we have shown, this conceptual modification works well in practice, so that one is 
able to reproduce quantum thermal averages through the annealing with different 
kick operators; on the contrary, using a single non-ergodic operators does not work.
However, a consequence of this modification is that now the annealing step,
or in other words the number of annealing steps $n_a$, is not only relevant
to the success probability (i.e., of finding an eigenstate of 
the Szegedy operator with eigenvalue 1) but also to the 
fact that the selected state is actually the CETS. In other words, 
there is a systematic effect in the algorithm, related to the fact that 
one might find the wrong state, and this systematic is a function of $n_a$.
In particular, we have shown that the error scales as $1/n_a$, at least for
$n_a$ large enough.

Going back to the comparison with the QMS, the final situation seems different from
initial expectations. In the QMS, the main algorithm-specific systematics is related to the number 
of rethermalization steps performed along the Markov chain: in this case, 
the systematic error is exponentially suppressed in such number. 
The outcome of our study is that the main algorithm-specific systematics of Q$^2$MA
scales as the inverse of the number of annealing steps, which is a less favorable 
polynomial scaling, compared to QMS.

Finally, we have explored the systematic effects related to the inexact representation
of the system energy spectrum obtained through the Quantum Phase Estimation (QPE), 
which is a problem affecting a large class of quantum algorithms and could be avoided 
for the particular explored system, due to its simplicity. 
In this case, the systematic error is expected to be suppressed as the number
of qubits used to represent the energy register is increased, i.e., as the grid
of possible outcomes of the QPE is made finer and finer. We have shown that this 
is indeed the case when the number of qubits is large enough, with a non-trivial 
intermediate regime due to the interplay between the energy level spacing, the grid spacing
and overall range explorable by QPE. As a final comment, one should consider
that this intermediate regime is likely to be less relevant for real and more complex many-body systems,
for which the distribution of energy levels is more chaotic.

Future developments along the same research line should consider and compare different approaches,
like those based on variational quantum algorithms or on quantum simulators~\cite{MariCarmen_rev_2020}, as well as less trivial models, including 
gauge degrees of freedom.

\acknowledgments

This study was carried out within the National Centre on HPC, Big Data and 
Quantum Computing - SPOKE 10 (Quantum Computing) and received funding from 
the European Union Next-GenerationEU - National Recovery and Resilience 
Plan (NRRP) – MISSION 4 COMPONENT 2, INVESTMENT N. 1.4 – CUP N. I53C22000690001. 
This manuscript reflects only the authors’ views and opinions, neither the 
European Union nor the European Commission can be considered responsible for them. 
It is a pleasure to acknowledge inspiring discussions with Raffaele Tripiccione, 
Leonardo Cosmai and Fabio Schifano during the early stages of this work. 
We thank Man-Hong~Yung and Al{\'a}n~Aspuru-Guzik for correspondence.
The work of Claudio Bonanno is supported by the Spanish Research Agency 
(Agencia Estatal de Investigaci\'on) through the grant IFT Centro 
de Excelencia Severo Ochoa CEX2020-001007-S and, partially, by grant PID2021-127526NB-I00, 
both funded by MCIN/AEI/ 10.13039/501100011033. Salvatore Tirone acknowledges support from projects PRIN 2017 Taming complexity via Quantum Strategies: a Hybrid Integrated Photonic approach (QUSHIP) Id. 2017SRN-BRK and PRO3 Quantum Pathfinder. The research of Kevin Zambello has been supported 
by the University of Pisa under the “PRA - Progetti di Ricerca di Ateneo” (Institutional
Research Grants) - Project No. PRA 2020-2021 92 “Quantum Computing, Technologies and 
Applications”. Numerical simulations have been performed at the IT Center 
of the University of Pisa and on the \texttt{Marconi100} machines at CINECA, based on 
the agreement between INFN and CINECA, under projects INF22\_npqcd and INF23\_npqcd.

\bibliographystyle{apsrev4-1}
\bibliography{refs}

\end{document}